\documentclass[twocolumn]{revtex4}
\usepackage{amsmath,dcolumn,feynmf,epsfig,CJK}
\begin{document}
%\begin{CJK*}{GBK}{song}
\title{POSSIBLE GENERATION OF A $\pi$-CONDENSATION IN FREE SPACE BY COLLISIONS BETWEEN PHOTONS AND PROTONS}
\author{QI-REN ZHANG }
\address{School of Physics, Peking University, Beijing 100871,
China}

%\date{}
%\parskip 0pt
%\parindent 5mm
\begin{abstract}  A sharply peaked structure is found in the angular distribution of emitted $\pi^+$  mesons from the photon-proton collisions. It offers a possible way for generating a $\pi^+$-condensation in free space. To make the stimulated emission of $\pi^+$-mesons efficient, a ring resonator is designed.

{\bf Key words:}\hskip 5mm Hadron quantum electrodynamics,Stimulated $\pi^+$-emission, $\pi$-condensation in a free space. Ring resonator.

{\bf PACS number(s)}: 13.60.Le, 03.70.+k, 03.75.Nt, 29.20.db
\end{abstract} \maketitle
%\end{CJK*}

\vskip0.3cm

\section{Introduction}
Possible $\pi$-condensation in nuclear matter was first considered in seventies last century\cite{sa,sc}.
To put the work on a better mathematic foundation we exactly solved the Dirac equation for the nucleon in
 a classical $\pi$-field\cite{z}.
A theory of nuclear matter was therefore proposed in the form of relativistic mean field theory, in which
a mean $\pi$-field is taken into account\cite{z1,z2,zg,g}. To minimize the energy per-nucleon, a nonzero value of
mean $\pi$-field appears. It shows the possible existence of the $\pi$-condensation in nuclear matter.
$\pi$-condensation is also considered in quark-gluon plasma\cite{ss,hjz,chh}. However,
until now we have not seen any direct experimental evidence for the existence of a $\pi$-condensation,
neither in nuclear matter nor in free space. It makes us eager to generate a $\pi$-condensation experimentally,
in free space first. Of course, the generation of a $\pi$-condensation in free space itself is interesting.

We designed a way for generating a
$\gamma$-ray laser using the photon-electron collisions\cite{z3,z4}. It is because we see that, the angular
distribution of the outgoing $\gamma$-photons is sharply peaked. Since the energy of an emitted $\gamma$-photon
is a definite function of its outgoing direction, the $\gamma$-photons are emitted almost into one state.
In this paper, we shall show that a similar situation appears for the $\pi^+$\!-meson emissions in the
photon-proton collisions. The angular distribution of the emitted $\pi^+$\!-mesons is also sharply peaked.
We may therefore design a similar way for generating a $\pi^+$-condensation in free space, by use of
photon-proton collisions.

\section{Angular distribution of the emitted $\pi^+$-mesons in a head on photon-proton collision\label{A2}}
Consider the reaction
\begin{equation}\gamma+{\rm p}\rightarrow\pi^++{\rm n},\label{1}\end{equation}
in which a $\gamma$-photon and a proton p head on collide with each other, and transit into a $\pi^+$-meson
and a neutron n.
When the energy of the photon is not too high, the colour
degrees of freedom in hadrons are not important. The problem may therefore be handled by hadron quantum electrodynamics.
Denote the proton field, neutron field, charged meson field and photon field by $\Psi_p \, , \Psi_n , \Phi $ and $(A_\mu)$
respectively, the Lagrangian density for the system is
\begin{eqnarray}{\cal L}={\cal L}_p+{\cal L}_n+{\cal L}_m+{\cal L}_\gamma+{\cal L}_s,\label{2}\end{eqnarray}in which
\begin{eqnarray}{\cal L}_p&=&-\overline{\Psi}_p[\gamma_\mu(\partial_\mu-{\rm i}eA_\mu)+m]\Psi_p\, ,\label{3}\\
{\cal L}_n&=&-\overline{\Psi}_n(\gamma_\mu\partial_\mu+m)\Psi_n\, ,\label{4}\\
{\cal L}_m&=&-[(\partial_\mu+{\rm i}eA_\mu)\Phi^*(\partial_\mu-{\rm i}eA_\mu)\Phi+m_\pi^2\Phi^*\Phi]\, ,\label{5}\\
{\cal L}_\gamma &=&-\frac{1}{4}(\partial_\mu A_\nu -\partial_\nu A_\mu )(\partial_\mu A_\nu -\partial_\nu A_\mu)
\, ,\label{6}\end{eqnarray}
and
\begin{eqnarray}\!\!\!\!\!\!\!\!\!\!\!\!\!\!\!\!\!\!\!\!\!\!\!\!\!\!
{\cal L}_s &=&{\rm i}\sqrt{2}G(\overline{\Psi}_p\gamma_5\Psi_n\Phi+\overline{\Psi}_n\gamma_5\Psi_p\Phi^*)\label{7}
\end{eqnarray}
are Lagrangian densities of protons with their electromagnetic
interactions, neutrons, $\pi^\pm$-mesons with their\,
electromagnetic \,interactions, photons, \,and the strong
interactions between related hadrons, respectively. The\! nature\!
unit \!system of $\hbar\!=c\!=\!1$ is used.  Symbols  are\! defined
in the usual way as given in standard \!text books, for examples in
\cite{l,we}. $e$ and $G$ are interaction constants for
electromagnetic interaction and strong interaction respectively,
with the corresponding values $\alpha\equiv e^2/4\pi=1/137.\cdots$
and $\alpha_s\equiv G^2/4\pi=14.6$.

For the reaction (\ref{1}), a factor of electromagnetic interaction appears always with a factor of strong interaction.
The constants $\alpha$ and $\alpha_s$ always appear together in the form of a product $\alpha\alpha_s$.
Since $\alpha\alpha_s<1$, a perturbation treatment for the reaction  seems reasonable. The lowest order
transition matrix element for the reaction is
\begin{eqnarray}&&\langle\sigma_n,\mbox{\boldmath$q$},\mbox{\boldmath$\kappa$}|T|\mbox{\boldmath$k$},
\mbox{\boldmath$e$},
\mbox{\boldmath$p$},\sigma_p \rangle=-Ge\frac{\prod_{\mu=0}^3\delta(p_\mu\!+\!k_\mu\!-\!q_\mu\!
-\!\kappa_\mu)}{\sqrt{2}(2\pi)^2\sqrt{k\kappa_0}} \nonumber\\ &&\times
\overline{u}_{\sigma_n}(\mbox{\boldmath$q$})\gamma_5\left[\frac{1}{{\rm i}\gamma^\mu(p_\mu
+k_\mu)+m}{\rm i}\gamma^\mu e_\mu \right. \nonumber\\&&\left.+\frac{1}{(p^\mu-q^\mu)(p_\mu-q_\mu)+m_\pi^2}2\kappa^\mu
e_\mu\right]u_{\sigma_p}(\mbox{\boldmath$p$}),\label{8}\end{eqnarray}
in which $[p_\mu],[q_\mu],[k_\mu],$ and $[\kappa_\mu]$, with $\mu=0,1,2, \mbox{and} \, 3,$  are energy-momentum four
vectors of proton, neutron, photon, and $\pi^+$-meson respectively. $[e_\mu]$ with $\mu=0,1,2, \mbox{and}\, 3,$ is the
polarization four vector of the photon. $m$ and $m_\pi$
are masses of the nucleon and the charged pion respectively. $u_{\sigma_n}(\mbox{\boldmath$q$})$ is the Dirac spinor of
the neutron with spin $\sigma_n$ and momentum $\mbox{\boldmath$q$}$, and $u_{\sigma_p}(\mbox{\boldmath$p$})$ is that of
the proton with spin $\sigma_p$ and momentum $\mbox{\boldmath$p$}$. $\delta-$functions in (\ref{8}) show energy-momentum
conservation in the reaction (\ref{1}).\! They, together with the energy-momentum relations
$p_0=\sqrt{p^2+m^2},q_0=\sqrt{q^2+m^2},\kappa_0=\sqrt{\kappa^2+m_\pi^2}$ and $k_0=k$, give
\begin{eqnarray}&&(A^2-1)\kappa^2+2AB\kappa+B^2-m_\pi^2=0,\label{81}\\
&&A=\frac{k-p}{p_0+k}\cos\theta\;\;\mbox{and}\;\;B=\frac{m_\pi^2+2k(p_0+p)}{2(p_0+k)},\label{82}
 \end{eqnarray}
 $\theta$ is the angle between moving directions of the incident photon and the emitted pion. $\kappa>0$ is the absolute
value of the pion momentum, therefore should be the positive root of
equation (\ref{81}). It defines the energy $\kappa_0$ as a function
of the moving direction $\theta$ for the pion.

Take the Coulomb gauge, in which the contribution from longitudinal and temporal components of the electromagnetic
field is collected in the coulomb energy between charged particles, and is negligible when space charge effect being
unimportant. Only the contribution from the transverse components of the electromagnetic field will be considered in the
following. Let $\mbox{\boldmath$e$}_{i}{\rm e}^{{\rm i}\mbox{\boldmath$k$}\cdot\mbox{\boldmath$x$}}$ with $i=1,2$ show
the transverse plane wave, we have $e_{i0}=0, \mbox{\boldmath$e$}_i\cdot\mbox{\boldmath$k$}=0$ for $i=1,2$,
and $\mbox{\boldmath$e$}_i^*\cdot\mbox{\boldmath$e$}_{i'}=\delta_{i£¬i'}$.

For experiments without measuring spins, all transition
probabilities and cross-sections have to be summed up over the final
spin states and averaged over the initial spin states. Using the
projection operator method, we obtain
\begin{widetext}
\begin{eqnarray}
\!\!\!\!\!\!\!\!\!\!\frac{1}{4}\!\sum_{i=1,2}\sum_{\sigma_n=-1,1}\sum_{\sigma_p=-1,1}\!\left|
\overline{u}_{\sigma_n}(\mbox{\boldmath$q$})\gamma_5\!\!\left[\frac{1}{{\rm i}\gamma^\mu(p_\mu
+k_\mu)+m}{\rm i}\gamma^\mu e_{i\mu} \!\!+\!\frac{1}{(p^\mu-q^\mu)(p_\mu-q_\mu)+m_\pi^2}2\kappa^\mu e_{i\mu}\!\right]
\!\!u_{\sigma_p}(\mbox{\boldmath$p$})\!\right|^2\!\!\!=\!\frac{X}{p_0q_0},\label{9}\end{eqnarray}
\begin{eqnarray}X&=&\frac{1}{2}f^2\left\{\left[2k(p+p_0)-\kappa_0(k+p_0)+\kappa(k-p)\cos\theta\right]
\left[k(2p+p_0)-p^2\right]+2pk(p_0+p)(p-k+\kappa\cos\theta)
\right.\nonumber\\&+&\left[p^2+p_0(k-\kappa_0)\right](p-k)^2-(p^2+p_0k)(p-k)(p-k+\kappa\cos\theta)\left.\right\}
+\kappa^2g(\theta)\sin^2\theta\left\{fk(p+p_0)\right.
\nonumber\\&+&g(\theta)\left[p_0(k-\kappa_0)+p(k-\kappa\cos\theta)\right]\left.\right\},
\label{10}
\end{eqnarray}\end{widetext}
with $f=1/[2k(p_0+p)]$, and $g(\theta)=1/[2k(\kappa\cos\theta-\kappa_0)]$.
The transition probability per-unit time for the pion goes into a differential solid angle ${\rm d}\Omega$ is
\begin{eqnarray}\frac{{\rm d}P}{{\rm d}t}=2\alpha_s\alpha \frac{XJ\kappa^2}{k\kappa_0q_0p_0}\,{\rm d}\Omega\,
{\rm d}\kappa\,\delta\!(E_f\!-\!E_i).\label{101}\end{eqnarray}$J=1/V$ is the incident photon current density in our unit
 system, and $V$ is the volume of the reaction space. $E_i=p_0+k$ and $E_f=q_0+\kappa_0$
are initial and final energies of the process respectively. Under fixed initial momenta $\mbox{\boldmath$p$}$ and
$\mbox{\boldmath$k$}$,
\begin{eqnarray}{\rm d}\kappa=\frac{\kappa_0(p_0+k-\kappa_0)}{(p_0+k)\kappa+\kappa_0(p-k)\cos\theta}{\rm d}(E_f-E_i).
\label{102}\end{eqnarray}
We therefore have
\begin{eqnarray}\frac{{\rm d}P}{{\rm d}t}=2\alpha_s\alpha XYJ\lambda_\pi^2{\rm d}\Omega .\label{11}\end{eqnarray}
 $\lambda_\pi$ is the Compton wavelength of the charged pion, and
\begin{eqnarray}Y=\frac{\kappa^2m_\pi^2}{kp_0[(p_0+k)\kappa+\kappa_0(p-k)\cos\theta]}.\label{12}\end{eqnarray}
The differential cross-section for the photo-production of the charged pion on a proton is therefore
\begin{eqnarray}\frac{{\rm d}\sigma}{{\rm d}\Omega}=2\alpha\alpha_sXY\lambda_\pi^2.\label{13}\end{eqnarray}
Notice, $X$ and $Y$ are dimensionless.
\begin{figure}\includegraphics[width=7cm]{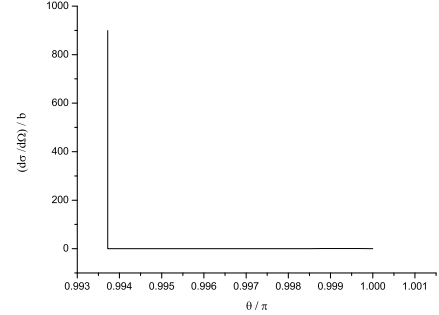}
\caption{Relation between the differential cross-section in unit of barn for pion emission and the angle
$\theta$ in unit of $\pi$, in a head on collision of an 1.4MeV photon and a 434GeV proton }\label{fig1}
\end{figure}\begin{figure}\includegraphics[width=7cm]{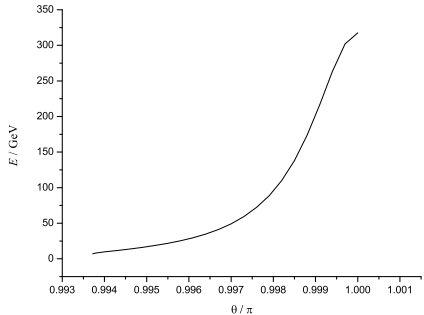}
\caption{Relation between the emitted pion energy $E$ in unit of GeV and the angle $\theta$ in unit of $\pi$,
in a head on collision of an 1.4MeV photon and a 434GeV proton}
\label{fig2}
\end{figure}

An example of numerical results is shown in figures \ref{fig1} and \ref{fig2}. The angular distribution shown in figure
\ref{fig1} is rather characteristic. It is sharply peaked, and is therefore favorable for emitting pions into the most
probable state.
However, the most probable state is not unique. In the example shown in figure \ref{fig1}, the most probable emission
directions distribute on the surface of a cone, each along a generatrix. The vertex of the cone is at the reaction point,
and the axis is on the incident line of the proton. The angle between the generatrix and the axis is 0.006278$\pi$.
In the following we shall show that a resonance mechanism makes almost all emitted pions go to one selected most
probable state, therefore generates a $\pi^+$-condensation in free space.

\section{Stimulated emission, resonance, and the ring resonator for the $\pi$-condensation\label{A3}}
The key ingredient for making a laser is the stimulated emission of radiation. This is also true for making a
$\pi$-condensation in the free space.
If there are already $N$ pions in a state, the transition probability for emitting one more pion into this state
has to be multiplied by an extra factor $N+1$. The equation (\ref{11}) is therefore generalized to
\begin{eqnarray}\frac{{\rm d}P}{{\rm d}t}=2(N+1)\alpha_s\alpha XYJ\lambda_\pi^2{\rm d}\Omega,\label{111}\end{eqnarray}
which includes contributions of\, both\, spontaneous \,and\, stimulated emissions of pions into a given state.
Here we see a positive feedback between  pion numbers of already in and emitted into a given state. The result is
a collapse of pion population into the most probable states.
\begin{figure}\includegraphics[width=7cm]{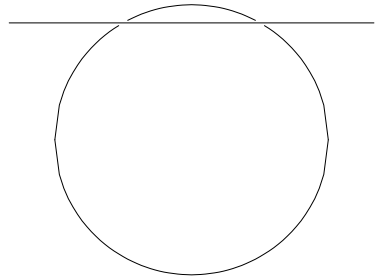}
\caption{A schematic diagram of the ring resonator for the stimulated emission of $\pi^+$-mesons in head on collisions
between photons and protons.}
\label{fig3}
\end{figure}

To make emitted pions condense in one state, we need a resonance mechanism.
Figure \ref{fig3} shows a schematic designation of a ring resonator for the stimulated emission of $\pi^+$-mesons in
head on collisions between photons and protons. The straight line denotes the incident line, and the circle denotes the
storage ring. They intersect each other at two gaps on the ring. Photons and protons inject from left and right
respectively along the incident line, and are designed to collide with each other at the gaps.  Elementary geometry
tells us that angles between tangents of the circle at two gaps and the incident line equal each other, so that we may
design these tangents along the most probable directions for pion emission at both gaps. $\pi^+$-mesons emitted on this
most probable direction therefore enter the inner space of the ring. It is its central circular channel. They move along
the channel under the interaction of an appropriate constant magnetic field perpendicular to the ring plane.
A coincidence of colliding photons and protons together with the earlier emitted $\pi^+$-mesons at the gaps is designed,
so that the stimulated emissions may happen. The already stored $\pi^+$-mesons stimulate the new $\pi^+$-meson emissions
at the gap. The $\pi^+$-mesons are emitted along the most probable direction selected by the ring, and enter the storage ring at the gap under the interaction of the magnetic field. They are therefore prepared to stimulate the next $\pi^+$-meson emissions at the next gap.  In this way a resonance is formed and a special most probable emission is selected at each gap.

In our example of head on collision between a 1.4MeV photon and a
434GeV proton, the energy of the most probable emitted $\pi^+$-meson
is 6.9GeV, as shown in figure \ref{fig2}. In a magnetic field of
$B=1$T, they would move in a storage ring of radius 23m. Using the
data shown at the end of last section, we see that two gaps on the
ring in figure \ref{fig3} open a central angle of $0.012556\pi$ at
the center of the ring. Therefore the length of the arc between
these two gaps on the ring of radius 23m is 0.9m.  One may therefore
open many pairs of gaps on the ring to intensify the condensed pion
beam many times in one circle. On the other hand, the half-life time
of a 6.9GeV $\pi^+$-meson is $8.95\times10^{-7}$s. A half of
$\pi^+$-mesons in the beam may move 268.33m before their decay. It
is about twice of the ring circumference. It seems, we may generate
a rather intense condensed pion beam in this way, and store it in
the ring. However, there are various interactions between
$\pi^+$-mesons. Among them, the long range electromagnetic
interaction may be important at not too high density of the meson
beam. This is the so called space charge problem. The electric force
of the space charge is outward perpendicular to the  meson beam, and
the magnetic force of the moving space charge is inward  perpendicular
to the  meson beam. They together may make nonzero probability for
mesons to leave from the resonance orbits, and limit the beam
density. It offers a saturation mechanism for the $\pi-$condensation
in our example. Fortunately, effects of these electric and magnetic forces
kill each other, especially for high velocity mesons. Their total
effect approaches zero at the relativity limit $v\rightarrow c$, and
therefore makes no serious problem against generating a $\pi^+$-condensation with an enough intensity.
Finally, charged pions running in a ring may lose energy by Bremsstrahlung. But it may be easily
compensated by usual acceleration techniques.
\section{Conclusions}
The $\pi$-condensation may be generated in a way like that in the laser generation. First of all, we need a pion source.
In our example proposed here, it is played by the hadronic reaction (\ref{1}).  Various sharply peaked spectra of pion
emission appear.  They make emissions concentrate to some specified pion states, and therefore are welcome.
We then need a way for realizing the stimulated pion emissions to start the $\pi$-condensation, and need a resonance
mechanism for selecting a special pion state to condense. These are designed in the ring resonator shown in the last
section.

The reaction (\ref{1}) was analysed by the hadron electrodynamics. Hadron-dynamics is not a fundamental theory, but
an effective theory only. Therefore we should not rely on its quantitative results. We may obtain the quantitative
results directly by experiments. However, some qualitative characters do not depend on the dynamical details.
In the derivation and the numerical calculation we see, that the sharply peaked structures of the angular
distribution for pion emissions are connected directly with the relativistic energy-momentum conservation
relations (\ref{81}) and (\ref{82}). It is a result of kinetics governing the reaction, and therefore is reliable.
The designation of the ring resonator is based on fundamental electromagnetism, and is therefore reliable too.
It makes us believe that our proposal is worthy to try experimentally.

\end{document}